\documentclass[pmlr,twocolumn,10pt, openany]{jmlr} 

\usepackage{booktabs}
\usepackage{siunitx}

\theorembodyfont{\upshape}
\theoremheaderfont{\scshape}
\theorempostheader{:}
\theoremsep{\newline}

\jmlryear{2022}
\jmlrworkshop{Conference on Health, Inference, and Learning (CHIL) 2022} 

\title[Short Title]{Full Title of Article\titlebreak This Title Has
A Line Break}

 \author{%
  \Name{Shalini Saini\nametag{\thanks{This work was done at the University of Alabama at Birmingham}}} \Email{s.saini@tamu.edu}\\
   \Name{Nitesh Saxena} \Email{nsaxena@tamu.edu}\\
  \addr Texas A\&M University, College Station, TX, USA
  }

\usepackage{url}
\usepackage{breakurl}

\newcommand{\quotes}[1]{``#1''}

\newcolumntype{P}[1]{>{\centering\arraybackslash}p{#1}}
\newcolumntype{R}[1]{>{\raggedleft\arraybackslash}p{#1}}                                                                                                                                               
\newcolumntype{L}[1]{>{\raggedright\arraybackslash}p{#1}}  

\title{Predatory Medicine: Exploring and Measuring the
			\titlebreak Vulnerability of Medical AI to Predatory Science}

\begin{document}

\maketitle

\begin{abstract}

Medical Artificial Intelligence (MedAI) for diagnosis, treatment options, and drug development represents the new age of healthcare. The security, integrity, and credibility of MedAI tools are paramount issues because human lives are at stake. MedAI solutions are often heavily dependent on scientific medical research literature as a primary data source that draws the attacker’s attention as a potential target. We present a first study of how the output of MedAI can be polluted with Predatory Publications Presence (PPP).

		We study two MedAI systems: \textit{mediKanren} (Disease independent) and \textit{CancerMine} (Disease-specific), which use research literature as primary data input from the research repository \textit{PubMed}, PubMed derived database \textit{SemMedDB}, and NIH translational Knowledge Graphs (KGs). Our study has a three-pronged focus: (1) identifying the PPP in PubMed; (2) verifying the PPP in SemMedDB and the KGs; (3) demonstrating the existing vulnerability of PPP traversing to the MedAI output. 
		
		Our contribution lies in identifying the existing PPP in the MedAI inputs and demonstrating how predatory science can jeopardize the credibility of MedAI solutions, making their real-life deployment questionable.

\end{abstract}

\paragraph*{Data and Code Availability}

This paper uses the data and code from \href{https://pubmed.ncbi.nlm.nih.gov/
}{\textit{PubMed}},  \href{https://lhncbc.nlm.nih.gov/ii/tools/SemRep_SemMedDB_SKR/SemMedDB_download.html}{\textit{NIH SemMedDB}}, NIH Translational KGs  (provided by mediKanren team), \href{https://zenodo.org/record/4304808#.X_cwITSSlPb}{\textit{CancerMine Zenodo repository}}, and \href{https://github.com/webyrd/mediKanren}{\textit{mediKanren Github}}

\section{Introduction}
\label{sec:intro}
The undeniable contribution of Artificial Intelligence (AI) in medicine is reaching new milestones. In 2018, FDA approved \textit{IDx-DR} as the first human-independent AI system to detect Diabetic Retinopathy, a leading cause of blindness \citep{fdaEye}. Google’s AI \textit{Inception v3} may assist with the early detection of lung cancer, the deadliest cancer, causing 1.7 million deaths per year globally \citep{googleBlogLungCancer}. 
 
\smallskip \noindent{\textbf{MedAI Relevance to Precision Medicine:}}
\textit{Precision Medicine} is an innovative approach based on an individual’s genetics, health history, environment, and diet. Biomedical research-driven Precision Medicine is promising to provide improved healthcare and lower the overall burden of unknown, delayed, or incorrect diagnosis and treatment \citep{singh2017global,tehrani201325,orphan_disease}. Precision Medicine relies heavily on data and analytics for its adoption into healthcare \citep{ginsburg2018precision}. MedAI assists Precision Medicine by processing a large amount of information from heterogeneous sources in no time. 

\smallskip \noindent {\textbf{Impact of Research on MedAI:}}
National Institute of Health (NIH) is the largest public investor of biomedical research globally, investing more than \$30 billion a year aiming to provide improved and affordable healthcare \citep{nih-research}. NIH-National Library of Medicine (NLM) maintains the comprehensive research publication source \textit{PubMed} comprising more than 32 million citations for biomedical literature \citep{pubmed}. Many MedAI systems rely upon scientific research publications in medicine as the primary data source for knowledge extraction.

If research gets manipulated through bogus, plagiarized, biased, or fraudulent conclusions, it turns into predatory research, potentially harming patients directly or indirectly. As per Wang et al., around 15\% of retraction in biomedical Open Access Journals is due to fraudulent data \citep{wang2019retracted}. Anesthesiologist Scott Reuben and Family Medicine practitioner Anne Kirkman Campbell are few among many others to commit such frauds for money, fame, and position, causing direct harm to patients including loss of lives \citep{Reuben, FDATrial}. Multiple studies about certain vaccines causing disorder in children baffled the medical community and the public for more than a decade \citep{rao2011mmr,wakefield1998retracted}. 

\smallskip \noindent{\textbf{Data Pollution in MedAI Solutions:}} 
MedAI solutions, which utilize research literature as primary data input, are prone to data pollution. Passive data pollution attacks and active adversarial attacks can impact the integrity and security of MedAI solutions. Untargeted predatory publications induce passive data pollution, while an active adversarial attack is a targeted approach with deliberately poisoning the publication databases through specific predatory journals. Targeted and untargeted predatory publications can inject new data to poison the input dataset. We focus on existing and presumably untargeted data pollution, which can potentially influence the output of the MedAI. 

\smallskip \noindent {\textbf{Our Contributions:}}
In this paper, we report on a novel case study of passive data pollution in MedAI solutions mediKanren \citep{medikanren}, and CancerMine \citep{lever2019cancermine} to verify the vulnerability of research publications. To the best of our knowledge, this is a first study to explore the existing data pollution and demonstrate the threat in real-world MedAI solutions.  

Our work casts serious doubt on whether research literature-based MedAI solutions are reliable enough to take chances with human lives. Moreover, our results show that MedAI may have no built-in logic to mitigate such threats.

\smallskip \noindent \textbf{Why is this a Security Study?} 
Predatory research infiltration in MedAI solutions and its impact on MedAIs’ output is an exploitable vulnerability, which is new to the security community. While our study is an interdisciplinary effort, we believe that the security community should firsthand know the threat of research literature-based data pollution impacting MedAI solutions. The current and future MedAI systems may avoid these pitfalls if aware of the threats.  We also contacted mediKanren team highlighting the underlying vulnerability.
\begin{figure*}[!htb]
	\centering
	\vspace{-1mm}
	\includegraphics[width=0.9\textwidth]{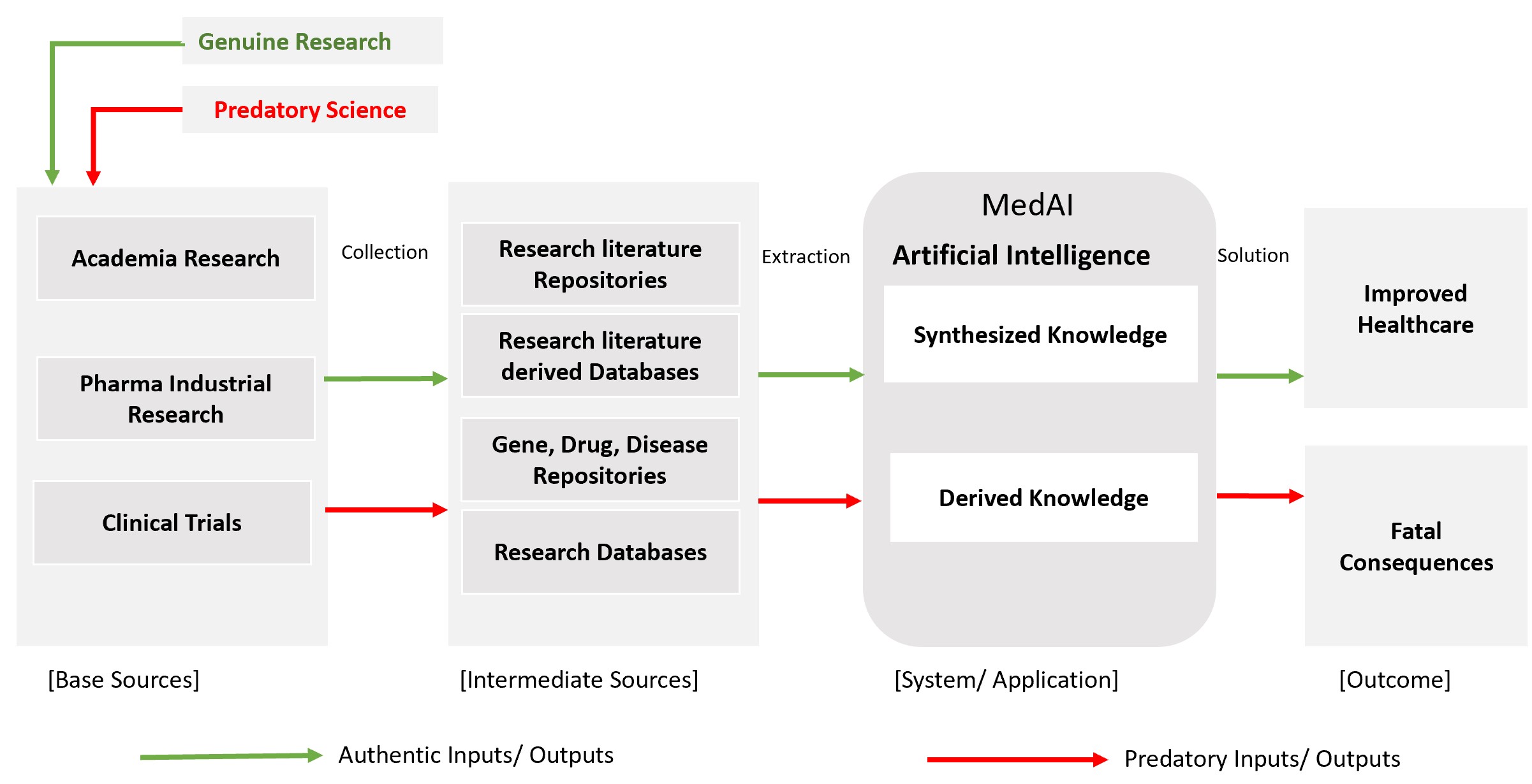}
	\vspace{0mm}
	\caption{High-Level Overview of MedAI Components and Involvement of Predatory Science}
	\label{fig:MedAI Overview}
\end{figure*}

\section{Background}
\label{sec:background}
``Artificial Intelligence" (AI) in medicine is being employed in robotic procedures, diagnostics, statistics, and human biology, including \textit{omics} \citep{hamet2017artificial}. Though it is still a far-fetched idea to replace the human touch in medicine, MedAI has opened possibilities to save on manual efforts and time to provide faster and adequate decision-making. 

\subsection{Health Care Revolution through MedAI}
Around one in 10 Americans is affected by some rare disease, and 80\% of around 7000 known rare diseases are genetics-based \citep{genetic_rare_disease}. Orphan disease diagnosis may take from 1 to 5 years \citep{orphan_disease} while rare diseases patients suffer from 40\% wrong initial diagnosis and 5 to 30 years wait for correct diagnosis \citep{faurisson2004survey}. 

A  2021 data shows that 230 Startups are using AI in drug discovery to improve success rate significantly \citep{drugai}. A recent effort reported finding successful novel FDA-approved therapeutic recommendations for disorders ranging from undiagnosed and purely symptomatic disease to genetically diagnosed metabolic disorders \citep{uab2seconds, medikanren}. In 2016, Wang et al. reported that pathology image-based MedAI could correctly identify metastatic breast cancer with 92\% accuracy. A human expert could determine with 96\% accuracy but applying both led to 99.5\% accuracy \citep{wang2016deep}. There is compelling evidence that MedAI can play a vital role in enhancing and complementing the `medical intelligence’ of the future clinician \citep{ramesh2004artificial,genetic_rare_disease}.

\subsection{Key Methodologies Adopted by MedAI}
With advances in computational power and big data, machine learning (ML) is the most widely used AI component in MedAI solutions. Deep Learning (DL) and Natural Language Processing (NLP) are widely employed AI methods to extract meaningful information from the research literature. Artificial Neural networks and fuzzy logic combined as a hybrid intelligent system can accommodate common sense, extract knowledge from raw data, and use human-like reasoning mechanisms \citep{nih_ml}. KGs are highly applicable in the medical domain and research as knowledge reasoning can find relationships among diagnosis, diseases, and treatments. Logical inference and probabilistic refinements can develop intelligent systems to suggest treatment options.

\subsection{Security and Integrity of MedAI} 
Modern Healthcare solutions using AI look promising to save time, money, effort significantly, but the cost of ``trusted but manipulated" information from such MedAI solutions is too high to ignore. Figure \ref{fig:MedAI Overview} depicts a high-level overview to show how predatory inputs can compromise a MedAI. An unreliable MedAI output can be fatal in clinical settings, and erroneous results can misalign the overall cycle of future research and healthcare solutions towards the harmful direction. Adversarial attacks on neural networks can cause errors in identifying cancer tumors and damage the confidence in MedAI output \citep{kotia2019risk}. Szegedy et al. showed that very subtle adversarial inputs, which may not appear as pathological, can potentially change the output \citep{leung2015machine}. A critical insight into why adversarial examples exist is that, given any dataset, the attacker can potentially perturb it in a direction that aligns well with the weights of the MedAI algorithm and thus amplifies its effect on the output \citep{goodfellow2014explaining}. As NLP is the core process to extract intelligent information from the research literature for MedAI, a robust and efficient algorithm against NLP adversarial attacks is a necessity rather than preference \citep{zhang2020adversarial}. MedAI aims to bring all the relevant data together and filter out irrelevant information without skipping more challenging instances. It is crucial for MedAI, especially in rare diseases, to include that one paper with the latest finding that may alter the life for some patient(s), and it is even more critical to validate if that paper is predatory or not.

A real challenge is maintaining the integrity and security of research data as the basis of MedAI in Precision Medicine and not letting it be Predatory Medicine. Our study highlights the vulnerability of polluted input through predatory publications that may have the potential to generate unreliable MedAI output, especially in finely targeted scenarios of Precision Medicine.

\label{sec:PredatoryResearch}

\section{Predatory Research}

Medical research has been revolutionary in the past few years, but innovation is not the only reason for the high volume of research publications. `Publish or Perish’ culture puts enormous pressure on budding scientists and researchers to publish their research \citep{rawat2014publish}. Publications and citations are being used as a metric for progressing towards the doctorate, employment, promotions, and grants/ funding by state and federal agencies. Opportunists may exploit these trends for their benefit to lure easy targets looking for some publication credits \citep{rawat2014publish}. Research misconduct is an even more significant threat to pollute the research repositories \citep{smith2006research, wang2017retraction,wang2019retracted}. Misleading conclusions may go undetected for an extended period and may affect clinical practices before being retracted \citep{budd1999effects}. Accommodating  information from heterogeneous sources is vital in MedAI decision-making; it also enables predatory research to get mixed with the authentic inputs.
 
\subsection{What is a Predatory Journal?} 
Currently, the characteristics of predatory journals have not been standardized nor broadly accepted by the research community. The majority of potential predatory journals have absent or minimal peer-review process \citep{bartholomew2014science,richtig2018problems} and follow questionable practices focusing on `pay to publish' model \citep{richtig2018problems}. With a possibility of having plagiarized, incorrect, fake data and manipulated results, predatory journals are, in fact, increasingly interfering with genuine research. Jeffrey Beall was first to raise the concern around 2008 and maintained a list of possibly predatory journals. DOAJ, Cabell’s list, and other independent online resources are reference points to identify potentially predatory journals \citep{predatoryjournals,grudniewicz2019predatory,doaj}. In the absence of any standard definition, we rely on the current list of potential predatory journals from these resources to apply in this work.

\subsection{How Big is Predatory Research?} 
Beall identified few potential predatory publishers in 2011, and by 2015, there were estimated as many as 10,000 predatory journals worldwide \citep{shen2015predatory}. The ultimate risk is the altered results of synthesized research because of rapidly increasing numbers of such predatory publications \citep{cobey2018predatory,bartholomew2014science}. There have been efforts to expose such practices of accepting fake papers, recruiting fake editors, but numbers are rising every year \citep{fakepapers,piotr_2017}. A genuine concern is that there are more suspected predatory journals (10,406) than legitimate journals (10,077) in Cabell’s list \citep{grudniewicz2019predatory}. 

\subsection{PPP Infiltration in Trusted Resources} 
The question is whether credited research resources like PubMed are already infected with predatory publications or not. Manca et al. reported the possible infiltration of predatory research in PubMed \citep{manca2018predatory}. European databases also carry predatory journals and research shows that predatory journals get even more publications than non-predatory journals after being listed in a reputed database \citep{perlin2018predatory}. 

\label{sec:MediKanren}

\section{Studied MedAI Solutions }
In this work, we focus on research literature based clinical MedAI solutions, which can impact patients directly. We study two MedAI solutions \textit{mediKanren} and \textit{CancerMine} with different AI approaches, data processing,  and output representation \citep{medikanren,lever2019cancermine}. Both studied MedAI solutions heavily rely on research literature inputs, primarily from PubMed. mediKanren has a broad scope of drug repurposing to treat newly diagnosed or unknown diseases based on inferred relationships. CancerMine aims to help with the early diagnosis of cancer type based on patients' genetic profiles. 
\begin{table*}[!h]
	\centering
	\scriptsize
	\vspace{2mm}
	\caption{Selected 25 Concepts, covering pandemics, common, and rare diseases, to analyze mediKanren prototype `code' vulnerability to PPP}

	\label{tab:25Concepts}
	\begin{tabular}{|L{4.3cm}|L{6.3cm}|L{1.1cm}|L{1.1cm}|}

		\hline
		\textbf{Concept   Name}           & \textbf{Query Term}                                                     & \textbf{Category} & \textbf{Type}    \\ \hline
		ADNP                              & ADNP, Helsmoortel-VanDerAa Syndrome,   HVDAS                            & very rare         & disease          \\ \hline
		Adenoid Cystic Carcinoma          & Adenoid Cystic Carcinoma, cylindroma                                    & rare              & disease          \\ \hline
		BCR                               & BCR                                                                     & common            & gene             \\ \hline
		Cervical Cancer                   & Cervical cancer                                                         & rare              & disease          \\ \hline
		Colorectal Cancer                 & Colorectal cancer, colon cancer,   rectal cancer                        & common            & disease          \\ \hline
		Coronavirus                       & Coronavirus  , HCoV-OC43, SARS-CoV-2                                    & pandemic          & virus            \\ \hline
		Curcumin                          & Curcumin, Diferuloylmethane, Turmeric                                   & common            & substance        \\ \hline
		Dravet Syndrome                   & Dravet Syndrome, Severe myclonic   epilepsy of infancy, SMEI            & very rare         & disease          \\ \hline
		Ebola                             & Ebola, rVSV-ZEBOV, ebola virus                                          & pandemic          & virus            \\ \hline
		Epithelial Mesenchymal Transition & Epithelial-mesenchymal transition,   epithelial mesenchymal transition, & common            & cell             \\ \hline
		Gastric Carcinoma                 & Gastric Carcinoma, Gastric Cancer,   Stomach Cancer                     & rare              & disease          \\ \hline
		Imatinib                          & Imatinib, Imatinib mesylate                                             & common            & drug             \\ \hline
		Ischemic Stroke                   & Ischemic stroke                                                         & common            & disease          \\ \hline
		Malaria                           & Malaria, Plasmodium berghei,   Plasmodium falciparum                    & rare              & disease          \\ \hline
		Methyltransferas                  & Methyltransferase                                                       & common            & disease          \\ \hline
		MIR-200                           & MIR-200, MicroRNA-200                                                   & common            & gene             \\ \hline
		Neuroblastoma                     & Neuroblastoma, mycn                                                     & very rare         & disease          \\ \hline
		Neuroendocrine Prostate Cancer    & neuroendocrine prostate cancer                                          & rare              & disease          \\ \hline
		Non-Small Cell Lung Cancer        & Non-small cell lung cancer, NSCLC                                       & common            & disease          \\ \hline
		Ovarian cancer                    & Ovarian cancer, germ cell cancer, germ   cell tumor                     & rare              & disease          \\ \hline
		Pancreatic Cancer                 & Pancreatic cancer, Pancreatic   neuroendocrine                          & rare              & disease          \\ \hline
		Prostate Cancer                   & Prostate cancer                                                         & common            & disease          \\ \hline
		T-Cell                            & T-Cell, t cells, Chimeric Antigen   Receptor                            & common            & cell             \\ \hline
		Triple Negative Breast Cancer     & Triple negative breast cancer                                           & rare   & disease          \\ \hline
		Tyrosine Kinase                   & Tyrosine , Tyrosine Kinase                                              & common            & Protein  \\ \hline
	\end{tabular}
	\end{table*}

\subsection{mediKanren} 
\textit{mediKanren} is a MedAI employing logical reasoning over the NIH SemMedDB and translational knowledge graphs to reduce the cost of drug discovery and repurposing \citep{uabMedikanren}. mediKanren is an implementation of miniKanren (Logic Programming Language), Racket (General-purpose Programming Language), Pubmed-derived SemMedDB, NIH translational KGs, and a graphical user interface (GUI) to simplify data exploration to assist Precision Medicine \citep{medikanren}. mediKanren also utilizes KGs with the standardized structure to improve interoperability, ingesting and processing new data from different sources. CURIE (Compact Uniform Resource Identifier), Concept Normalization, and KGs are the basis of mediKanren's functionality. 

We study state-of-the-art mediKanren prototypes, \textit{code} and \textit{biolink} \citep{githubMediKanren}. mediKanren is currently under real-world stress-testing, which makes it relevant to study for the existing exploitable vulnerabilities \citep{uab2seconds}.

\subsection{CancerMine} 
\textit{CancerMine} is an automated text-mining approach for extracting gene-disease relationships from PubMed literature to reduce the manual effort and cost of providing timely diagnosis and treatment \citep{singhal2016text}. CancerMine provides a database of drivers, oncogenes, and tumor-suppressors for different types of cancer. CancerMine extracts information from the PubMed, PubMed Central Open Access (PMCOA) subsets, and PubMed Central Author Manuscript Collection (PMCAMC). CancerMine uses the supervised machine learning approach using the Logistic Regression classifier on word frequencies and semantic features \citep{lever2019cancermine}. CancerMine is currently incorporating information from 35,623 PubMed publications. We study the publicly available CancerMine data downloaded in January 2021 to verify how predatory research is navigating to the output of this ML-based MedAI. Without exploiting any ML logic, the focus of this work is to identify and verify the existence of predatory research in the MedAI output.

\label{sec:experiments}
\section{Preliminaries: Resources and Setup}
The study environment was built on Linux OS using Singularity container 2.6.1 utilizing High-Performance Cluster (HPC). We reconstructed the NIH SemMedDB tables for mediKanran analysis using Open GPLv2 MariaDB ver 10.3.10 MySQL. CancerMine inputs and outputs in TSV/CSV file formats were analyzed using MS Excel.
\subsection{Key Terms and Definitions} 
Each article on PubMed has a unique identifier called \textit{PMID} (PubMed ID). In this work, we are using the SemMedDB (semmedVER40\_R), with PubMed data processed up to June 30, 2018. 

NIH defines a rare disease as a condition that affects fewer than 200,000 people in the US. A rare disease is known as an orphan disease if drug companies are not interested in developing treatments \citep{genetic_rare_disease, orphan_disease}. \textit{Concept} is defined as a unique medical term as per NIH Unified Medical Language System (UMLS) Metathesaurus, and \textit{CUI} is the Concept Unique Identifier \citep{bodenreider2004unified}. \textit{PREDICATE} is the representation of the relationship between any two medical concepts identified as SUBJECT and OBJECT. For example, Imatinib (SUBJECT) Treats (PREDICATE) Mastocytosis (OBJECT). 
Compact Uniform Resource Identifiers (CURIEs) serve as machine-readable markers for different databases. Graph vertices represent medical concepts in the KGs, while directed graph edges depict relationships between concepts. Edges also show metadata about source and evidence backing the represented relationship \citep{medikanren}. 


\section{Data Extraction}
We extracted a set of predatory journals based on Beall’s list\footnote{\href{https://beallslist.net/}{\textit{Beall's list of potential predatory journals and publishers}}.} and compared with more current list\footnote{\href{https://web.archive.org/web/20211220083526/https://predatoryjournals.com/journals/}{\textit{Web Archives-predatoryjournals.com}}.}. We created an advanced query with the list of existing predatory journals in PubMed, and downloaded the metadata for 47,051 predatory publications. We extracted 8,289 retracted publications from PubMed with \quotes{Retraction of publication [Publication Type]},  a query adopted from \cite{budd1999effects}.

\subsection{mediKanren Data Inputs}
We study 25 concepts to study in this work under common, rare, very rare, and pandemic categories to see the extent of PPP in diverse scenarios. For each of these 25 concepts, we extracted the set of 200 rows (100 rows of predatory PMIDs and 100 rows of non-predatory PMIDs) from the PREDICATION table of SemMedDB. A .csv file was prepared for each concept except for a few rare concepts with less than 200 publications. Table \ref{tab:25Concepts} provides details of the selected concepts.

Prototype biolink is currently utilizing data inputs from four NIH translational knowledge graphs \textit{RoboKop}, \textit{SemMed}, \textit{Orange}, and \textit{Rtx}. mediKanren team provided the local copy of NIH KGs employed in biolink. We applied python scripts to extract PMIDs from applied NIH Translational KGs to study the presence of predatory PMIDs. 

\subsection{CancerMine Data Inputs}
We utilized data from \href{https://zenodo.org/record/4304808#.X_cwITSSlPb}{\textit{CancerMine Zenodo repository}}. The primary raw input cancermine\_unfiltered.tsv is processed to create other two main inputs as cancermine\_collated.tsv and cancermine\_sentences.tsv. cancermine\_collated.tsv contains the cancer gene roles supporting citation counts, and cancermine\_sentences.tsv carries the sentences for the cancer gene roles. This SENTENCE file contains information on the source publication (e.g., journal, publication date, etc.), the actual sentence, and the cancer type, gene, and role extracted from the sentence.

\section {Passive Data Pollution Verification}
Based on our hypothesis, PubMed has existing data pollution with the predatory publications, and MedAI solutions use this polluted dataset. Other than predatory journal publications, the PubMed-derived database also carries the retracted research publications. To verify PPP in SemMedDB tables, we queried restructured SemMedDB on HPC to find predatory PMIDs. 

For mediKanren prototype `code', we run each concept-specific CSV file through Racket commands to pre-process the input. We execute all the end-user queries on Racket ver 7.4 and gui-simple.rkt file as working GUI for code prototype of mediKanren. `Biolink' prototype works with gui\_simple\_v3.rkt GUI on KGs, and we run sample queries to verify the data pollution in the input and output of the studied MedAI solutions.

For CancerMine, we verified the overall PPP in provided datasets from their repositories, queried online tool, and downloaded data to verify the PPP in tool-provided subsets. We utilized the same set of predatory PMIDs extracted from PubMed for mediKanren and CancerMine PPP verification.

\label{sec:results}

\section{Results}
Our results show that predatory publications have a significant presence in the research literature repository PubMed (47,051 predatory publications). Results also validate that predatory research can traverse from PubMed to MedAI output. For detailed analysis, we organize our results into five sub-sections: (1) PPP in SemMedDB; (2) Retracted publications in SemMedDB; (3) PPP in NIH Translational KGs; (4) PPP in mediKanren GUI output; and (5) PPP in CancerMine.

\subsection{PPP in SemMedDB}
The PREDICATION table is carrying concept, sentence ID, predicate, SUBJECT\_CUI, and OBJECT\_CUI. Although predatory PMIDs numbers seem smaller compared to the huge dataset size, the fraction holds a large number (44,665) of predatory publications. Table \ref{tab:PPP_SemMed} shows the overall PPP infiltration in SemMedDB tables.  

\begin{table}[!htb]
	\centering
	\scriptsize
	\vspace{-2mm}
	\caption{PPP in SemMedDB Tables}
		\vspace{2mm}
	\label{tab:PPP_SemMed}
	\begin{tabular}{|L{2cm}|R{1.7cm}|R{1.7cm}|}
		\hline
		\textbf{SemMed Table} & \textbf{Total  Rowcount} & \textbf{PPP  Rowcount} \\ \hline\hline
		PREDICATION      & 97,772,561        & 256,641         \\ \hline
		SENTENCE       & 187,449,479        & 357,384         \\ \hline
		ENTITY        & 1,369,837,426       & 2,735,289        \\ \hline
		CITATION       & 29,137,782        & 44,670         \\ \hline
	\end{tabular}
	\end{table}
\begin{table}[!htb]
	\centering
	\scriptsize
	\vspace{-2mm}
	\caption{SemMedDB: PPP For Studied 25 Concepts}
	\vspace{2mm}
	\label{tab:specific25ppp}
	\begin{tabular}{|L{4cm}|L{1.2cm}|R{1cm}|}
		\hline
		\textbf{Concept}      & \textbf{Category}   & \textbf{PPP}  \\ \hline
		ADNP                    & very rare & 5 \\ \hline
		Adenoid Cystic Carcinoma         & rare & 51 \\ \hline
		BCR                    & commom & 659 \\ \hline
		Cervical Cancer            & rare  & 1,005\\ \hline
		Colorectal Cancer            & common & 5,510\\ \hline
		Coronavirus                 & pandemic & 6\\ \hline
		Curcumin                 & common & 500\\ \hline
		Dravet Syndrome              & very rare & 1 \\ \hline
		Ebola                    & pandemic & 15\\ \hline
		Epithelial Mesenchymal Transition     & common & 679\\ \hline
		Gastric Carcinoma            & rare & 3,727\\ \hline
		Imatinib                  & common & 424\\ \hline
		Ischemic Stroke              & common & 266 \\ \hline
		Malaria                   & rare. & 252\\ \hline
		Methyltransferase             & common & 57 \\ \hline
		MIR-200                  & common & 300 \\ \hline
		Neuroblastoma              & very rare & 1,269 \\ \hline
		Neuroendocrine Prostate Cancer       & rare  & 11\\ \hline
		Non-small cell Lung Cancer       & common & 4,143 \\ \hline
		Ovarian Cancer              & rare & 2,718\\ \hline
		Pancreatic  Cancer           & rare & 2,139 \\ \hline
		Prostate  cancer            & common & 3,849 \\ \hline
		T-Cell                  & common & 4,378\\ \hline
		Triple  Negative Breast Cancer       & rare & 246\\ \hline
		Tyrosine  Kinase            & common & 2,061 \\ \hline
	\end{tabular}
	
\end{table}

We also verified the PPP in SemMedDB for all 25 selected concepts in this work. PPP count varies from under 5 (very rare diseases: Dravet Syndrome, ADNP) to around 5000 or more(common cancer: Non-small-cell Lung Cancer, Colorectal Cancer). We can see that PPP is higher, in general, for common diseases, and lower for rare and very rare diseases, but also higher PPP for rare diseases like \textit{Gastric Carcinoma}. We did not analyze the percentage of concept-specific PPP for these 25 concepts as any presence may be damaging for a specific scenario. Table \ref{tab:specific25ppp} shows the PPP count in SemMedDB for all 25 concepts studied in this work.

\subsection{Retracted Publications in SemMedDB}
There are 8,289 retracted publications found on PubMed. An increase from 311 in 2010 to 860 in 2019 indicates the rising problem of retracted publications in PubMed. These retracted articles also exist in SemMedDB with their original PMIDs and with retraction notice PMIDs. A query with \quotes{Retracted:} returns 538 rows from the SENTENCE table, and the PREDICATION table returns 398 rows. 
  \begin{figure*}[!htb]
	\centering
	\includegraphics[width=.78\textwidth]{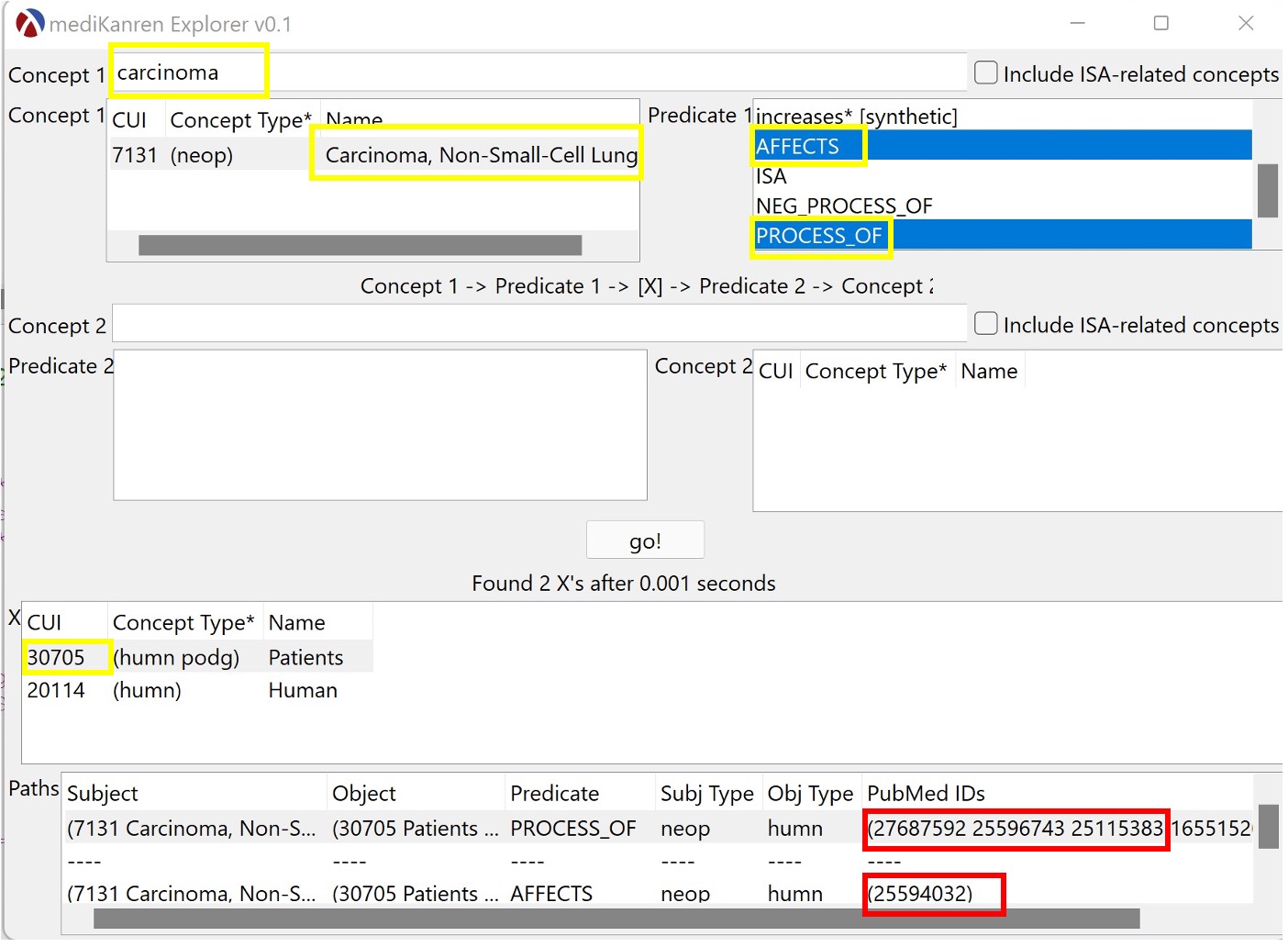}

	\vspace{-2mm}
	\caption{mediKanren `code' GUI [Concept, Predicate and Object- yellow; Predatory PMIDs - red]}
	\label{fig:MediKanren GUI}
\end{figure*}
\begin{table}[!htb]
	\centering
	\scriptsize
	\vspace{-2mm}
	\caption{PPP in NIH Translational KGs}
	\vspace{2mm}
	\label{tab:PPP_TranslatorDB}
	\begin{tabular}{|L{1.5cm}|R{1.7cm}|R{1.7cm}|}
		\hline
		\textbf{Database} & \textbf{Total Rowcount} & \textbf{PPP Rowcount} \\ \hline\hline
		Robokop              & 5,367,905      & 5,676          \\ \hline
		SemMed              & 44,245,576      & 164,324         \\ \hline
		Orange              & 836,118       & 287           \\ \hline
		Rtx                & 7,664        & 99            \\ \hline
	\end{tabular}
		
\end{table}

\subsection{PPP in NIH Translational KGs} 
mediKanren prototype `biolink' can create and run queries on translational KGs to deliver the suggestions in a comprehensible manner \citep{uabMedikanren}. Table \ref{tab:PPP_TranslatorDB} shows the PPP in all four translational KGs.  Percentage of PPP in these KGs does not appear huge, but PPP is significant, especially in the largest KG SemMed. For example, KG Rtx is with the most significant percentage of 1.3 with a smaller PPP (99). KG SemMed shows a smaller percentage (.37) with a much higher PPP (164,324). This observation also validates that even a lower number with a high percentage on any concept will have a higher probability to appear in MedAI output.
\begin{table*}[!htb]
	\centering
	\scriptsize
	\vspace{7mm}
	\caption{PPP in MediKanren `code' GUI Output with Predatory and Non-predatory PMIDs}	

	\label{tab:specificconceptGui}
	\vspace{4mm}
	\begin{tabular}{|L{4.3cm}|L{5.3cm}|P{.5cm}|P{1.1cm}|P{.6cm}|}
		\hline
		\textbf{Concept Name} &
		\textbf{(Subject\_CUI, Oject\_CUI, Predicate)} &
		
		\textbf{Pred} &
		\textbf{Non-Pred} &
		\textbf{Total } \\ \hline\hline
		 ADNP                                &  (1334473, 3394, INTERACTS\_WITH)   & 1 & 0 & 1     \\ \hline
		 Adenoid Cystic Carcinoma          &  (10606, 30705, PROCESS\_OF)        & 3 & 2 & 5      \\ \hline
		 BCR                                 &  (812385, 935989, COEXISTS\_WITH)   & 1 & 0 & 1     \\ \hline
		 Cervical Cancer                     &  (7847, 30705, PROCESS\_OF)         & 4 & 6 & 10     \\ \hline
		 Colorectal Cancer                   &  (7102, 27651, AFFECTS)             & 1 & 0 & 1     \\ \hline
		 Coronavirus                         &  (10076, 15576,  PROCESS\_OF)        & 1 & 0 & 1     \\ \hline
		 Curcumin                            &  (10467, 6826, AFFECTS)             & 1 & 0 & 1     \\ \hline
		 Dravet Syndrome                     &  (3064, 15827, LOCATION\_OF)        & 1 & 0 & 1     \\ \hline
		 Ebola                               & (282687, 30705, PROCESS\_OF)       & 1 & 3 & 4      \\ \hline
		 Epithelial Mesenchymal Transition & (14609, 346109, LOCATION\_OF)      & 1 & 0 & 1     \\ \hline
		 Gastric Carcinoma                   &  (24623, 30705, PROCESS\_OF)        & 2 & 5 & 7   \\ \hline
		 Imatinib                            & (935989, 23437, TREATS)            & 5 & 3 & 8    \\ \hline
		 Ischemic Stroke                     &  (948008, 30705, PROCESS\_OF)       & 7 & 4 & 11  \\ \hline
		 Malaria                             & (24530, 21311, ISA)                & 1 & 2 & 3   \\ \hline
		 Methyltransferas                    &  (25831, 11315, PART\_OF)           & 1 & 0 & 1     \\ \hline
		 MIR-200                             &  (1537839, 13081, AFFECTS)          & 1 & 0 & 1     \\ \hline
		 Neuroblastoma                       &  (27819, 27651, ISA)                & 2 & 2 & 4      \\ \hline
		 Neuroendocrine Prostate Cancer    &  (936223, 30705, PROCESS\_OF)       & 1 & 1 & 2      \\ \hline
		 Non-Small Cell Lung Cancer        &  (7131, 30705, NEG\_PROCESS\_OF)      & 7 & 4 & 11  \\ \hline
		 Ovarian cancer                      &  (29925, 1520166, PROCESS\_OF)      & 1 & 1 & 2      \\ \hline
		 Pancreatic Cancer                   &  (235974, 30705, PROCESS\_OF)       & 2 & 2 & 4      \\ \hline
		 Prostate Cancer                     &  (376358, 30705, PROCESS\_OF)       & 5 & 6 & 11  \\ \hline
		 T-Cell                              &  (279592, 30705, PROCESS\_OF)       & 1 & 0 & 1     \\ \hline
		 Triple Negative Breast   Cancer     &  (6142, 30705, PROCESS\_OF)           & 3 & 3 & 6      \\ \hline
		 Tyrosine Kinase                     &  (206364, 27651, ASSOCIATED\_WITH)  & 1 & 0 & 1     \\ \hline
	\end{tabular}
	\end{table*}

\subsection{PPP in mediKanren GUI Output}
Specific input datasets for 25 concepts are fed independently to mediKanren prototype \textit{code} to analyze targeted queries to follow the Precision Medicine approach. 
\begin{figure*}[!htb]
	\centering
	\vspace{0mm}
	\includegraphics[width=.77\textwidth]{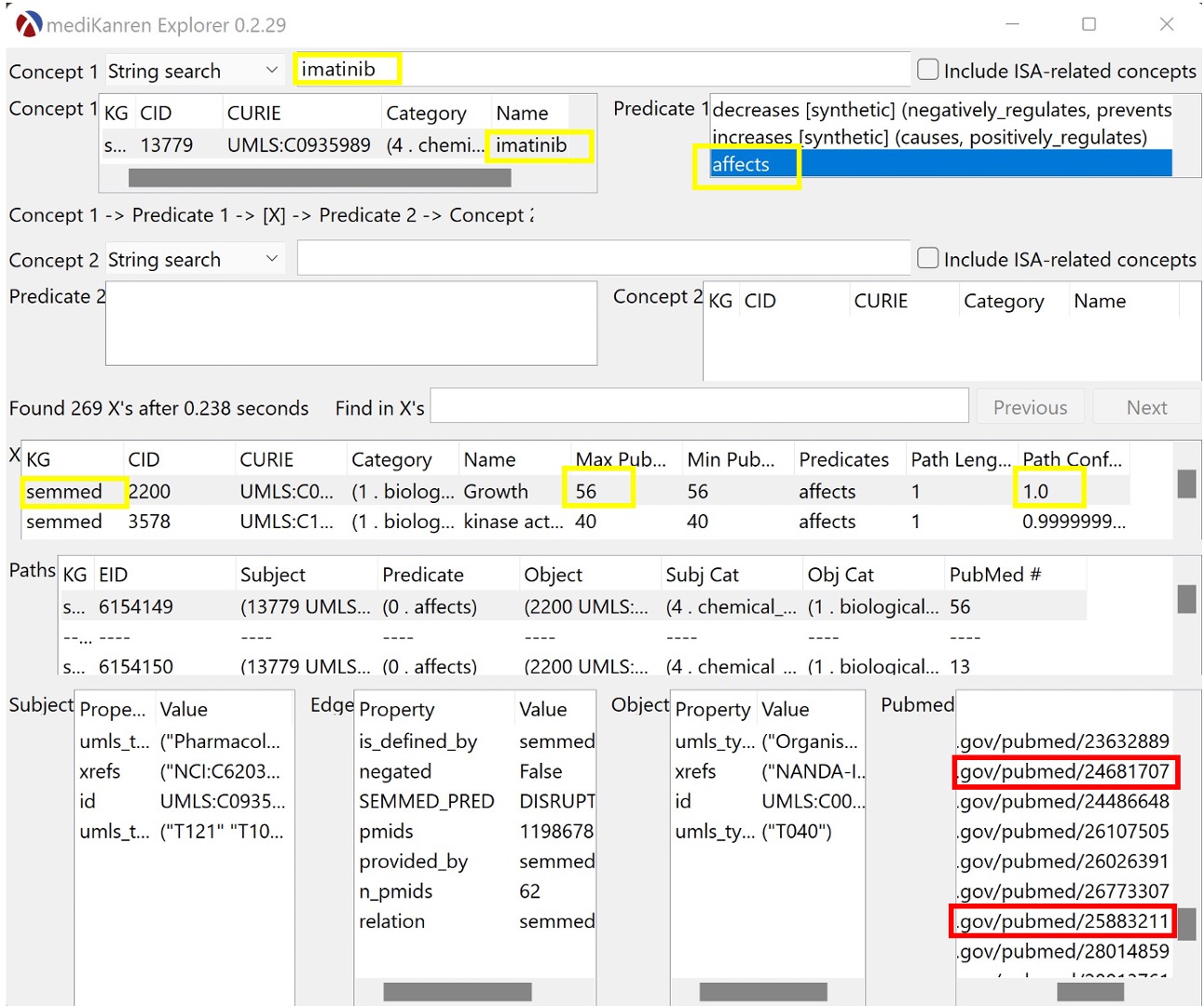}

	\vspace{2mm}
	\caption{mediKanren prototype `biolink' GUI [Concept, Predicate and Object, PubMed articles, and Path confidence- yellow; Some of Predatory PMIDs - red] }
	
	\label{fig:MediKanrenBiolink GUI}
\end{figure*}
We executed multiple queries on each concept for different triples of Subject\_CUI, Object\_CUI, and Predicate. We found that majority of queries show predatory PMID(s) in mediKanren GUI output. Figure \ref{fig:MediKanren GUI} shows non-predatory PMIDs appearing in mediKanren prototype \textit{code}. For example, the only PMID returned for concept \textit{Non-small-cell lung cancer} and predicate \textit{AFFECTS} is predatory PMID, and for predicate \textit{PROCESS\_OF} output contains both predatory and non-predatory PMIDs. 

Table \ref{tab:specificconceptGui} shows 25 representative cases for 25 concepts validating the problem of PPP traversing from PubMed to MedAI output. Data demonstrates that mediKanren may pick predatory PMIDs for common as well as for rare diseases if predatory PMIDs are present in the inputs. 

PPP in SemMedDB can affect the MedAI outcome even with a low number of predatory PMIDs. If a particular case has only a few publications on a specific concept, and there are more predatory than non-predatory, that may mislead the decision. For example, the concept \textit{Imatinib} for predicate \textit{TREATS} returns five predatory and three non-predatory PMIDs from a pool of 200 rows (100 predatory PMIDs and 100 non-predatory PMIDs). MediKanren GUI showed clear evidence of picking up predatory PMIDs in output with targeted input datasets. 

For mediKanren prototype \textit{biolink}, we queried the whole input for all four KGs without any specific test sampling. Figure \ref{fig:MediKanrenBiolink GUI} shows that predatory/ non-predatory PMIDs are appearing in prototype biolink’s output. 

We did not verify the presence of retracted PMIDs in mediKanren GUI output in this work. However, based on the working logic, MedAI is expected to pick any present publication (predatory or non-predatory) for a specific user query.

\subsection{PPP in CancerMine}
\begin{table}[!hb]
	\centering
	\scriptsize
	\caption{PPP in CancerMine Dataset}
	\vspace{2mm}
	\label{tab:cancermineall}
\begin{tabular}{|L{2cm}|R{2cm}|R{2cm}|}
	\hline
	\textbf{} & \textbf{unfiltered tsv} & \textbf{sentences tsv} \\ \hline
	Total PMIDs     & 172,443               & 55,881               \\ \hline
	PPP         & 9,479                & 2,873               \\ \hline
	PPP \%        & 5.50\%               & 5.14\%               \\ \hline
	Unique  PMIDs    & 73,099               & 35,623               \\ \hline
	Unique  PPP     & 3,384                & 1,637               \\ \hline
	PPP \%        & 4.63\%               & 4.60\%               \\ \hline
\end{tabular}
\end{table}

We analyze CancerMine data directly downloaded from the data repository and downloaded output files from online queries on the CancerMine web tool. Table \ref{tab:cancermineall} shows that overall predatory PMIDs have above 5\% in raw data and extracted Sentence CSV file for cancer type and gene relationship. We observe that 51.64\% of predatory PMIDs have a prediction probability of 0.7 and higher (max is .9997), indicating the role of these publications on overall prediction probability. We Downloaded raw data from the CancerMine repository to carry 455 cancer types, and 150 out of 455 types show predatory PMIDs. Breast cancer shows the highest number of predatory PMIDs.  

\begin{table*}[!ht]
	\centering
	\scriptsize
	\vspace{0mm}
	\caption{CancerMine Predatory and Non-Predatory PMIDs for Cancer-Role-Gene Triple}
	\vspace{2mm}
	\label{tab:gencancerPred}
	\begin{tabular}{|L{3.5cm}|L{2.5cm}|L{1.5cm}|R{1cm}|R{1cm}|R{1cm}|}
		\hline
		\textbf{Type of Cancer}  & \textbf{Role}          & \textbf{Gene} & \textbf{PMIDs} & \textbf{Non-Pred} & \textbf{Pred} \\ \hline
		Glioblastoma       & Oncogene            & CDK4     & 5         & 4         & 1       \\ \hline
		Lung small cell cancer & Driver             & ALK      & 1         & 0         & 1       \\ \hline
		Colorectal cancer     & Oncogene            & KRAS     & 167        & 158        & 9       \\ \hline
		Prostate cancer      & Tumor Suppressor & PTEN     & 137        & 128        & 9       \\ \hline
		Breast cancer       & Oncogene            & FOXM1     & 8         & 7         & 1       \\ \hline
		Pancreatic cancer     & Driver             & KRAS     & 171        & 160        & 11      \\ \hline
		Neuroblastoma       & Driver             & MYCN     & 164        & 131        & 33      \\ \hline
		Stomach cancer      & Tumor Suppressor & TFF1     & 16         & 12        & 4       \\ \hline
		Pituitary cancer     & Oncogene            & PTTG1     & 17         & 15        & 2       \\ \hline
		Ovarian cancer      & Tumor Suppressor & BRCA1     & 123        & 119        & 4       \\ \hline
	\end{tabular}
	\end{table*}

We cross-referenced the specific gene-cancer-role details from the CancerMine data repository and application output downloads. We observe PPP in multiple subsets of targeted queries regarding a particular cancer type, gene, and role of the gene. Table \ref{tab:gencancerPred} is presenting PPP for few other cancer-gene-role triples of selected cancer types. We selected these cancer types similar to some concepts explored for other MedAI \textit{mediKanren} in this work. PPP can be much higher on a particular cancer-role-gene triple. For example, Neuroblastoma-Driver-MYCN filters a set of 164 PMIDs, and 20.12\% (33) PMIDs are predatory. Another case of Stomach cancer-Tumor\_Suppressor-TFF1 is a much smaller set of 16 PMIDS with 25\% (4) predatory PMIDs.

We successfully show these predatory PMIDs in the input sources (SemMedDB and CancerMine datasets) and the output (mediKanren GUI output and CancerMine website downloaded output), which verifies the existing threat in real-life  research-literature based MedAI solutions. Though we are not exploring the threat of ML manipulation in this work, it may be possibility to analyze predatory publications’ impact on shifting prediction probability.

\label{sec:Limitations and Challenges}
\section{Limitations and Challenges}
Though our work clearly shows the infiltration of PPP in PubMed and affecting the output of mediKanren GUI, there are several factors to limit the outcome of this study. One of the challenges is determining the valid list of all known predatory journals at any given point in time. 

The frequency of derived database updates also can affect the PPP. All the mediKanren queries for the code prototype are executed with targeted much smaller datasets for selected concepts to highlight the threat. We believe that results will be comparable even if we execute queries on the whole dataset. 

This work only focuses on the existing PPP verification without any current analysis of data training and prediction probability affected by PPP.

\label{sec:discussion}
\section{Discussion and Future Work}

As observed in our study, MedAI relies on the credibility of the reputed data source. There is no in-built defense logic in both the studied MedAI solutions to minimize the threat of predatory research influencing the output. Our work indicates the need to have a better defense system at the source level to minimize the threat of data pollution. 
 
We shared our findings and concerns with the mediKanren team. The mediKanren team has acknowledged the threat and is looking into the issue. Though this work is specific to research literature-based clinical MedAI solutions, we looked at other current state-of-the-art solutions for existing defense strategies towards predatory science, if any exists. Iris.ai is specific for researchers to provide relevant research literature based on the research hypothesis. We observed predatory publications in iris.ai results as well, and the iris.ai team also confirmed not to have any current defense mechanism to filter out potential predatory research \citep{iris}.

Though currently not explored but in a possible adversarial attack scenario, in near future, it can be viable to produce targeted predatory publications in bulk through new-age NLP text-generator tools like GPT-3 \citep{brown2020language, gpt3}. A targeted adversarial attack on a rare disease can make it look valid while injecting fake data with alternative conclusions. This approach may impact rare and unknown diseases scenarios more as there is little available research to cross-validate. 

Approximately 50\% medical providers showed concern about producing fatal errors and technical/operational glitches. These concerns resonate with the known ethical and regulatory challenges with MedAI solutions involving privacy, data integrity, accessibility, accountability, transparency, and liability \citep{intelAI, vayena2018machine}.

In the age of social media and web-based information, the future trustworthy information extraction will have more significant challenges and higher stakes. Future work will further study the impact of predatory research, including retracted publication on MedAI solutions. We plan to look at other security vulnerabilities to better defend against information pollution. 


\label{sec:conclusion}
\section{Conclusion}
Our work concludes that polluted inputs can cause a possible failure of any MedAI to deliver the intended output. Predatory research is on the rise and may further degrade reserach literature based MedAI solutions' credibility. Studied MedAI solutions treat all research data as trusted and do not consider predatory research-induced data pollution. In the absence of any defense, MedAI solutions may produce unreliable output. Existing data pollution fuels motivation for more targeting attacks.

Our study shows clear evidence of how predatory research information is navigating through publication channels and eventually may alter patient care decisions if used in the clinical settings without resolving the existing threat of predatory research intrusion. We are confident that verifying the vulnerabilities early in the process will contribute in developing more robust solutions for taking Precision Medicine to the next level in broader settings.

\section*{Institutional Review Board (IRB)}
This research did not require IRB approval.

\acks{We thank Matthew Might and Michael J Patton from the mediKanren team for assisting with a local copy of translational datasets and useful insights.}

\bibliography{references}

\end{document}